\documentstyle[aps,amsmath,epsfig,preprint]{revtex}

\begin{document}

\title{\large \bf NANOTCAD2D: Two-dimensional code for the simulation of
nanoelectronic devices and structures}
\author{\normalsize G. Curatola$^1$, G. Iannaccone$^2$}
\address{Dipartimento di Ingegneria dell'Informazione,
Universit\'a degli studi di Pisa, \\
Via Diotisalvi 2,
I-56122 Pisa, Italy\\
e-mail: {\tt $^1$g.curatola@iet.unipi.it;  $^2$g.iannaccone@iet.unipi.it}}

\maketitle
\begin{abstract}

In this paper we present NANOTCAD2D, a code for the simulation of the
electrical properties of semiconductor-based nanoelectronic devices
and structures in two-dimensional domains.  Such code is based on the
solution of the Poisson/Schr\"odinger equation with density functional
theory and of the continuity equation of the ballistic current.
NANOTCAD2D can be applied to structures fabricated on III-IV,
strained-silicon and silicon-germanium heterostructures, CMOS
structures, and can easily be extended to new materials. In
particular, in the case of SiGe heterostructures, it computes the
effects of strain on the energy band profiles.  The effects of
interface states at the air/semiconductor interfaces, particularly
significant in the case of devices obtained by selective etching, are
also properly taken into account.

\end{abstract}

\section{Program summary}
~\\ {\em Title of program}: NANOTCAD2D \\ {\em Program summary
URL:}:http://www.phantomshub.com \\ {\em Program available at:}
URL:http://www.phantomshub.com. (can be freely used via a web
interface) \\ {\em Computer on which the program has been tested:}
Linux PCs.  \\ {\em Programming Language used:} FORTRAN 77.  \\ {\em
Memory used to execute with typical data:} 50-100 Mb. \\ {\em
Keywords:} density functional theory, quantum simulation, ballistic
transport, nanoscale devices, nanoelectronics, nanotechnology computer
aided design.  \\ {\em Field of application:} simulation of nanoscale
semiconductor devices and structures.  \\ {\em Method of solution:}
Self-consistent solution of the Poisson-Schr\"odinger equation with
density functional theory, in the approximation of local density, and
of the continuity equation for the ballistic current with the
Newton-Raphson algorithm.  \\ {\em Restrictions:} The present
simulation tool works only with rectangular grids.

\section{Introduction}

The electrical properties of nanoscale semiconductor devices and
structures are typically determined by quantum confinement, that
strongly affects the density of states of electrons and holes, and, by
ballistic transport, that takes place if device length is smaller than
the scattering length.

An efficient simulation tool, aimed at understanding device behavior
and at designing optimized structures, must therefore include such
aspects, and achieve at the same time a reasonable trade-off between
efficiency and accuracy.

Here we present NANOTCAD2D, a program for the simulation of
nanoelectronic semiconductor devices, based on the self-consistent
solution of Poisson/Schr\"{o}dinger equation and of the continuity
equation for electrons and holes in the case of ballistic
transport. The program allows to consider quantum confinement in one
or two dimensions, and to address more complex structures that can be
divided into regions in which different types of confinement are
present.

NANOTCAD2D is oriented at two main classes of devices: $i)$ quantum
wires, i.e., structures with translation symmetry in one direction,
that are completely described by the geometry of the cross section;
and $ii)$ ballistic field effect devices, again with translation
symmetry in a direction perpendicular to that of electron motion.

Ballistic transport is simply included by assuming that the occupation
factor of states injected from a given reservoir is that of the
reservoir itself, i.e., obeys Fermi-Dirac distribution with the
reservoir's Fermi energy.

In addition, the code implements a simplified model for localized
states at the semiconductor surface exposed to air. Indeed, the
depletion of a one-Dimensional or a two-Dimensional Electron Gas (1DEG
or 2DEG) due to acceptor-like surface states is of primary importance
in determining the electrical properties of narrow quantum wires or
structures with large exposed surfaces.

At present, the code allows to consider common semiconductor
materials, such as silicon, AlGaAs, InGaAs, strained silicon and
silicon germanium. In the case of silicon germanium, in particular, we
have developed a procedure for taking into account the effect of
strain caused by different lattice constants on the band structure.

The paper has the following structure: in section III we describe the
physical model implemented in the code; in section IV we discuss the
numerical aspects algorithms. In section V simulation examples are
presented and in section VI the conclusions. The appendix contains the
user's manual.

\section{Physical Model}

The density of states for electron and holes depends on the degree of
quantum confinement assumed in the different device regions.  There
are three possibilities: If quantum confinement is strong in both
directions, the density of states is obtained by solving the
two-dimensional Schr\"odinger equation.  If quantum confinement is
strong only in one direction (say, $x$) the density of states is
written as a sum of two-dimensional subbands, the edges of which are
obtained by solving the Schr\"odinger equation in the $x$ direction
for each mesh point along the $y$ axis.  If confinement can be
considered very weak, the density of states of the bulk material is
used.

For materials with degenerate or quasi-degenerate minima in the
conduction band, such as for example silicon, that has six degenerated
minima, the density of states is computed with the effective mass
approximation for each minimum, taking into account mass
anisotropy. The same procedure is applied to degeneracy of valence
band maxima.

In the following, we will discuss in some detail the expressions for
the density of states and carrier density considering only one band
valley, for simplicity of notation. Extension to multiple valleys is
straightforward.

\subsection{Charge density for two-dimensional quantum confinement(1DEG-1DHG)}

Let us consider a region where the confinement for electrons is strong
in both $x$ and $y$ directions. In such a case, the local density of
states per unit of volume and energy near a conduction band minimum is
given by:
\begin{eqnarray}
N_{1D}(E,x,y) & = &
\frac{\sqrt{2 m_z}}{\pi \hbar} \nonumber \\
& \times &
\sum_{i}\left|{\Psi_{i}(x,y)}\right|^2
(E-E_{i})^{-\frac{1}{2}} u(E-E_{i}),
\end{eqnarray}
where $u(E-E_{i})$ is the Heavyside function, $\Psi_i$ is the solution
of the Schr\"odinger equation in two dimensions, i.e.,
\begin{equation}
- \frac{\hbar^{2}}{2m_{x}}
\frac{\partial^{2}}{\partial x^{2}} \Psi_i
- \frac{\hbar^{2}}{2m_{y}}
\frac{\partial^{2}}{\partial y^{2}} \Psi_i
+ E_c(x,y) \Psi_i = E_{i} \Psi_i
,
\label{eq1}
\end{equation}
$E_i$ is the corresponding eigenvalue, $E_c$ is the conduction band,
$m_s$, $s=x,y,z$ is the effective mass in the direction denoted by the
pedix, $\hbar$ is the reduced Planck's constant.  At this point, by
integrating the density of states multiplied by the Fermi-Dirac
occupation factor, the quantum electron density can be expressed as:
\begin{equation}
n= \frac{\sqrt{2 m_z k_BT}}{\pi \hbar} \sum_{i}
 \left|{\Psi_{i}}\right|^2
 F_{-\frac{1}{2}}\left(\frac{E_f-E_{i}}{k_BT}\right) \label{ncar2d}
\end{equation}
where $F_{-1/2}$ is the Fermi-Dirac integral of order $-1/2$ and $E_f$
is the Fermi energy~\cite{FermiIntegr}.

In order to compute the hole concentration we have to solve
Schr\"{o}dinger equation for heavy holes and for light
holes. Therefore, the conduction band in (\ref{eq1}) is substituted by
the inverted valence band $- E_v(x,y)$ and the eigenvalues $-E^h_i$,
are obtained.

Therefore the hole concentration $p$ becomes:
\begin{equation}
p= \frac{\sqrt{2 m^h_z k_BT}}{\pi \hbar}
   \sum_{i}\left|{\Psi_{i}}\right|^2
   F_{-\frac{1}{2}}\left(\frac{E^h_{i}-E_f}{k_BT} \right),
   \label{pcar2d}
\end{equation}
where $m^h_z$ is the effective mass for holes in the $z$ direction.

\subsection{One-dimensional quantum confinement: two-dimensional
electron or hole gas (2DEG-2DHG)}

In the case of strong confinement in only one direction (for example
along the $x$-direction), we assume that the density of states can be
decomposed in a quantum term along the confined direction ($x$) and a
semiclassical term in the other directions. The one-dimensional
Schr\"{o}dinger equation for electrons in the $x$ direction for a mesh
point $y$ can be written as:

\begin{equation}
- \frac{\hbar^{2}}{2m_{x}} \frac{\partial^{2}}{\partial x^{2}} \Psi_i
+ E_c(x,y) \Psi_i = E_{i}(y) \Psi_i
\end{equation}

As a consequence, the available states for electrons are grouped into
two-dimensional subbands and the density of states can be expressed as
follows:
\begin{equation}
N_{2D}(E,x,y)=\frac{\sqrt{m_{y} m_{z}}}{2 \pi \hbar^{2}} \sum_{i}
\left|\Psi_{i}(x,y)\right|^{2} u(E-E_{i}(y))
\end{equation}
The electron density therefore is:
\begin{equation}
n= \frac{k_BT\sqrt{ m_y m_z}}{\pi \hbar^2}
    \sum_{i}\left|{\Psi_{i}}\right|^2
    \ln\left[1+\exp\left(\frac{E_f-E_{i}}{k_BT}\right)\right] \label{ncar1d}
\end{equation}

Similar considerations apply to holes.

\subsection{Effects of surface states}

In the simulation of narrow semiconductor devices obtained with
selective etching, the effects of states at the exposed surface are
very important and must be taken into account in order to reproduce
with accuracy the experimental results. In particular, these states
can act as donors or acceptors and hence deeply affect the carrier
distribution within the device.

In order to correctly model the phenomenon, we have used a simple
model based on two parameters that is typically applied to
metal-semiconductor contacts \cite{Sze} and has been recently
validated for air-semiconductor interfaces \cite{pala}.

The two parameters are the density of interface states per unit energy
per unit area $D_S$ [eV$^{-1}$ cm$^{-2}$] and the energy difference
$\Phi^*$ between the vacuum level $E_o$ and the Fermi energy that
ensures a neutral charge at the interface.  States with energy below
$E_o - \Phi^*$ are donors and states with higher energy are acceptors.

Surface charge per unit surface can then be expressed as
$Q_s=-qD_S\left[E_f-\left(E_o-\Phi^*\right)\right]$, where $-q$ is the
electron charge.

\subsection{Poisson Equation}

All charge concentrations considered represent the source term of the
two-dimensional Poisson equation:
\begin{eqnarray}
&\nabla \cdot(\epsilon\nabla\Phi)=-\rho[\Phi] \nonumber \\ &
=-q\left[-n[\Phi]+p[\Phi]+N_D^+[\Phi]-N_A^-[\Phi]+\rho_s[\phi]\right]
\end{eqnarray}
where $\epsilon$ is the dielectric constant, $q$ is the electron
charge, $\rho_s$ is the term of surface charge per unit volume,
$N_D^+$ and $N_A^-$ the ionized donor and acceptor concentrations,
respectively~\cite{Sze}. Energy bands depend on the potential as:
\begin{eqnarray}
E_c(x,y) &=& E_c(x,y)|_{\Phi=0} -q\Phi(x,y) \nonumber \\
E_v(x,y) &=& E_v(x,y)|_{\Phi=0} -q\Phi(x,y)
.\end{eqnarray}

Potential and charge density profiles in equilibrium are
computed by solving the set of non linear partial differential
equations described above. The case of ballistic transport
is examined in the next section.

\subsection{Ballistic transport for electrons and holes.}

When carriers are injected into a semiconductor device, they are
likely to be scattered by a number of possible sources, including
acoustic and optical phonons, ionized impurities, defects, interfaces
and other carriers. If, however, device length is smaller than the
mean free path it is very likely for carriers to traverse the device
without suffering scattering events. Our code includes this type of
"ballistic" transport.

Ballistic transport is implemented here only in the case of
one-dimensional quantum confinement, when a description in terms of
two dimensional subbands is used. Indeed, we have shown that such
assumption involves a negligible error also in devices with channel
length of 25 nm \cite{fiorinanotech}.

\begin{figure}
\begin{center}
\epsfig{file=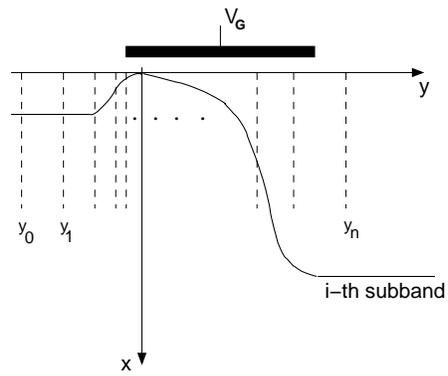,width=6cm}
\caption{Energy profile along the channel for a nanoscale FET. Only
electrons with energy higher than the barrier peak contribute
to the current.}
\label{mos}
\end{center}
\end{figure}

Let us consider Figure~\ref{mos}, representing a subband profile for
electrons along the channel. Carriers are injected into the channel
from a reservoir (source) and contribute to the current only if they
overcome the barrier modulated by the gate voltage and, to a lesser
degree, by the drain voltage (DIBL).

If we neglect the interaction between electrons and ions and among
electrons, we can simply assume that electrons with injected
longitudinal energy lower than the subband maximum are reflected back
to their originating contact, while the others are transmitted over
the barrier and contribute to the current~\cite{Natori,Lundstrom}.

Therefore, for each subband we evaluate the subband maximum $E_{imax}$
and the corresponding longitudinal position $y_{max}$.  All electrons
with longitudinal energy lower than $E_{imax}$ are in equilibrium with
the originating contact, while electrons with longitudinal energy
higher than $E_{imax}$ conserve the chemical potential of the
injecting reservoir. The occupation factor $f$ is therefore:
\begin{equation}
f(E,E_{F}) = \left\{
\begin{array}{ll}
\left[ 1 + \exp{\left(\frac{E-E_{FS}}{k_{B}T} \right)} \right]^{-1} &
{\rm if }\;\; y < y_{max}, \;\;E < E_{imax}, \\ \left[1 +
\exp{\left(\frac{E-E_{FD}}{k_{B}T} \right)} \right]^{-1} & {\rm if
}\;\; y > y_{max},\;\; E < E_{imax} \\ \left[1 +
\exp{\left(\frac{E-E_{FS}}{k_{B}T} \right)} \right]^{-1} + \left[1 +
\exp{\left(\frac{E-E_{FD}}{k_{B}T} \right)} \right]^{-1} & {\rm if
}\;\; E > E_{imax}
\end{array}
\right.
\end{equation}
where $E_{FS}$ ($E_{FD}$) is the source (drain) Fermi energy.  If we
write the total energy $E$ as $E=E_y+E_z$, where the term
$E_y=E_i+\frac{\hbar^2k^2_y}{2m_y}$ is the longitudinal energy, and
$E_z = \hbar^2 k^2_z/2m_z$ the transverse energy, the density of
states reads:
\begin{equation}
N_{2D}(E_y,E_z)\;\; dE=2\;\sum_i
|\Psi_i|^2\frac{\sqrt{m_y m_z}}{h^2}\frac{1}{\sqrt{E_y E_z}}\;\;dE_y
\;dE_z
\end{equation}
and the electron density is accordingly given by:
\begin{equation}
n=\int_0^{E_{imax}-E_i(y)} \sum_i |\Psi_i|^2 2\;\frac{\sqrt{m_y m_z}}{h^2}
{E_y}^{-\frac{1}{2}} \; dE_y \int_0^\infty {E_z}^{-\frac{1}{2}}
f(E,E_F) \; dE_z
\end{equation}

The electron current is evaluated assuming that there is no tunnel
current through the barrier so that only the electrons with
longitudinal energy higher than $E_{imax}$ can contribute.

\begin{eqnarray}
J_n = \int_0^\infty dE_z \int_{E_{imax}-E_i(y)}^\infty \sum_i \;\;
2\frac{\sqrt{m_y m_z}}{h^2} \frac{1}{\sqrt{E_yE_z}}
\sqrt{\frac{2E_y}{m_y}} \nonumber\\
\times \left[\frac{1}{1+exp\left(\frac{E_y+E_z-E_{FS}}{k_B
T}\right)}-\frac{1}{1+exp\left(\frac{E_y+E_z-E_{FD}}{k_B
T}\right)}\right] \;\; dE_y
\end{eqnarray}

Similar considerations apply to holes.

\section{Numerical Aspects}

\begin{figure}
\begin{center}
\epsfig{file=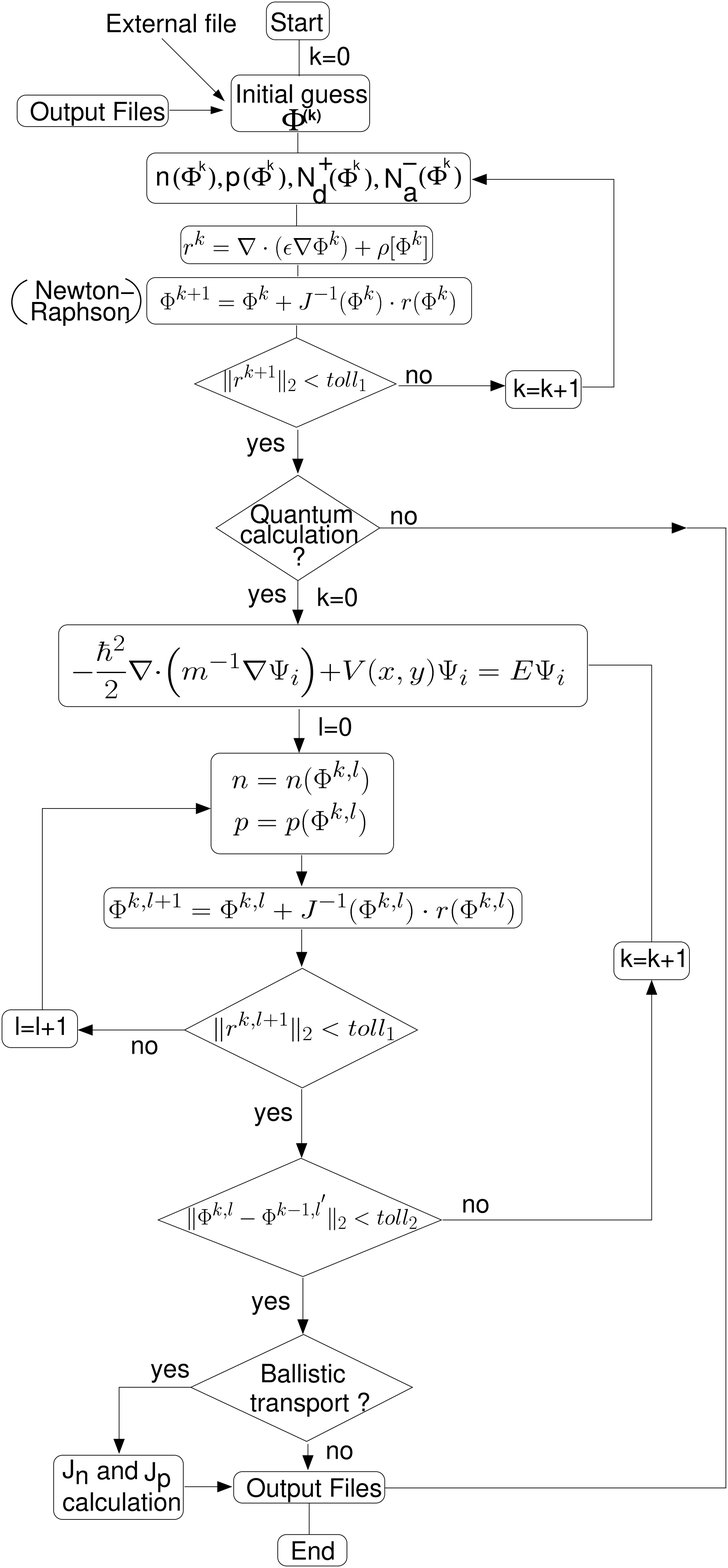,width=8cm}
\caption{Flow diagram of the algorithm implemented.} \label{flow}
\end{center}
\end{figure}

The flow diagram of the algorithm implemented is shown in
Figure~\ref{flow}. The program, using an initial guess for the
potential, starts with a semiclassical solution of the Poisson
equation. The equation is discretized with the Box-integration method
and solved with the Newton-Raphson algorithm.  Dirichlet boundary
conditions are enforced on each metal gate and homogeneous Neumann
conditions on the rest of the domain boundary. The solution obtained
is used as an initial guess for the quantum calculation, where the
nested Poisson/Schr\"odinger equation must be solved.

In order not to degrade convergence speed of the algorithm when also
the Schr\"odinger equation has to be solved, we have implemented a
simplified version of the predictor-corrector scheme proposed in
Ref.~\cite{Trellakis}. In this way, instead of solving both equations
at each Newton-Raphson step, we evaluate eigenfunctions and
eigenvalues only at the beginning of a Newton-Raphson cycle: for the
whole cycle eigenfunctions and the difference between the eigenvalues
and the energy bands in each point of the domain are assumed to be
constant. When a Newton-Raphson cycle ends the Schr\"{o}dinger
equation is solved again and a new cycle is started. The program ends
when the difference between the two-norm of the potential at the end
of two successive Newton-Raphson cycles is lower then a fixed
tolerance.~\cite{Ravaioli}

\begin{figure}
\begin{center}
\epsfig{file=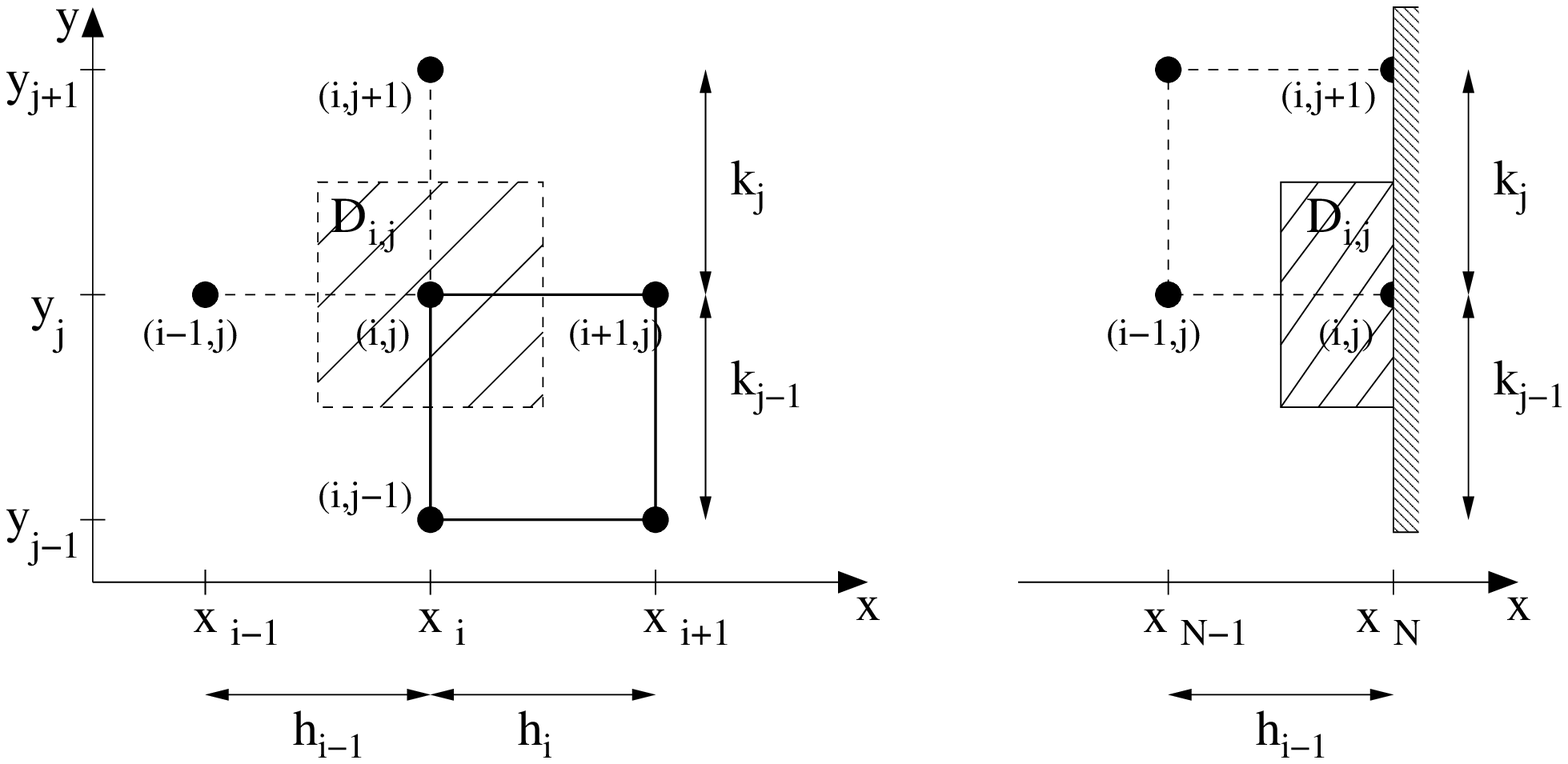,width=8cm}
\caption{$D_{i,j}$ represents the region associated to the grid point
$(i,j)$. In figure is presented the case of an internal point (left)
and a point in the edges (right).} \label{domain}
\end{center}
\end{figure}

In Figure~\ref{domain}(left) a rectangular uniform mesh is shown,
where it is possible to notice to the subdomain $D_{i,j}$ (dashed
line) associated to the generic grid point $(i,j)$. In our code, we
have distinguished the properties of the structure in {\em point}
characteristics and {\em material} characteristics.

Point characteristics are, for example, the potential, the Fermi
level, charge concentrations, while material properties (e.g., energy
gap, electron affinity, etc.) belong to the second group. In the
figure, is also shown (solid line) the material element connected to
the generic $(i,j)$ grid point.

In each subdomain, Poisson and Schr\"odinger equations are discretized
with box integration \cite{selbeherrer}

\subsection{Poisson Equation}

In particular, for the Poisson equation, after
integrating both members over the domain $D_{i,j}$, we obtain:
\begin{equation}
\int\int_{D_{i,j}}\nabla
\cdot(\epsilon\nabla\Phi(x,y))dxdy=-\int\int_{D_{i,j}}\rho(x,y)dxdy
\label{gilby}
,\end{equation}
which is discretized as
\begin{eqnarray}
\frac{\Phi_{i+1,j}-\Phi_{i,j}}{h_i}\left[\epsilon_{i,j+1}\frac{k_j}{2}+\epsilon_{i,j}\frac{k_{j-1}}{2}
\right]+ \nonumber \\
\frac{\Phi_{i,j+1}-\Phi_{i,j}}{k_j}\left[\epsilon_{i,j+1}\frac{h_i}{2}+\epsilon_{i-1,j+1}\frac{h_{i-1}}{2}
\right]- \nonumber \\
\frac{\Phi_{i,j}-\Phi_{i-1,j}}{h_{i-1}}\left[\epsilon_{i-1,j+1}\frac{k_j}{2}+\epsilon_{i-1,j}\frac{k_{j-1}}{2}
\right]- \nonumber \\
\frac{\Phi_{i,j}-\Phi_{i,j-1}}{k_{j-1}}\left[\epsilon_{i,j}\frac{h_i}{2}+\epsilon_{i-1,j}\frac{h_{i-1}}{2}
\right]= \nonumber \\
\left[-q\left(p_{i,j}-n_{i,j}+N_{Di,j}^+-N_{Ai,j}^-
\right)+\rho_{fi,j}\right]\frac{(h_i+h_{i-1})(k_j+k_{j-1})}{4}
\end{eqnarray}
where $h_i=x_{i+1}-x_i$, $k_j=y_{j+1}-y_j$, and pedices $i,j$ denote
the quantity in position $(x_i,y_j)$.

Proper boundary conditions must be enforced to the equation, such as
Dirichlet boundary conditions on each metal gate and homogeneous
Neumann conditions on the rest of the domain boundary. In the first
case, the potential $\Phi$ is fixed, while in the second case, the
electric field $\vec{\nabla}\Phi\cdot\vec{n}=-\varepsilon$ is fixed.

In the simulation code, we have chosen as reference level for energies
the vacuum level $E_o$ and hence the potential for each gate point is
obtained as:
\begin{equation}
\Phi(i,j)=E_o-\phi_{work}^n-E_F^{n}
\end{equation}
where $E_F^n$ and $\phi_{work}^n$ represent the Fermi level and the
work function of the n-th gate, respectively.

Let us consider a point $(i,j)$ on the boundary and let us make
reference to the Figure~\ref{domain}(right): observe how the region
$D_{i,j}$ is much smaller with respect to the case of internal point
($D_{i,j}$ becomes a quarter in the case of each of four vertexes grid
points). In this point, we enforce Neumann condition and the
discretization of the Poisson equation becomes:

\begin{eqnarray}
\varepsilon\left[\epsilon_{i,j+1}\frac{k_j}{2}+\epsilon_{i,j}\frac{k_{j-1}}{2}
\right]+ \nonumber \\
\frac{\Phi_{i,j+1}-\Phi_{i,j}}{k_j}\left[\epsilon_{i-1,j+1}\frac{h_{i-1}}{2}
\right]- \nonumber \\
\frac{\Phi_{i,j}-\Phi_{i-1,j}}{h_{i-1}}\left[\epsilon_{i-1,j+1}\frac{k_j}{2}+\epsilon_{i-1,j}\frac{k_{j-1}}{2}
\right]- \nonumber \\
\frac{\Phi_{i,j}-\Phi_{i,j-1}}{k_{j-1}}\left[\epsilon_{i-1,j}\frac{h_{i-1}}{2}
\right]= \nonumber \\
\left[-q\left(p_{i,j}-n_{i,j}+N_{Di,j}^+-N_{Ai,j}^-
\right)+\rho_{fi,j}\right]\frac{(h_{i-1})(k_j+k_{j-1})}{4}
\end{eqnarray}

\subsection{Schr\"{o}dinger Equation}

The two-dimensional single-particle Schr\"{o}dinger equation for
electrons, given a conduction band profile $E_c(x,y)$, reads:
\begin{equation}
- \frac{\hbar^{2}}{2} \nabla \cdot \left[m^{-1} \nabla \Psi_n\right]
  +E_c(x,y) \Psi_n =E_n \Psi_n
\end{equation}
where $\Psi_n(x,y)$ represents the $n$-th eigenfunction, $E_n$ is the
$n$-th eigenenergy, $m$ is the electron effective mass tensor in the
plane perpendicular to the direction of propagation,
\begin{equation}
m=\left[
\begin{array}{cc}
m_x & 0 \\
0 & m_y
\end{array} \right].
\end{equation}
In our simulations, we have discarded the exchange-correlation term,
since it provides a very small contribution.  Dirichlet boundary
conditions are enforced on the quantum simulation
domain.~\cite{pala_quantum}

With the box integration method, we obtain:

\begin{eqnarray}
-\frac{\hbar^2}{4}\left[\frac{\Psi_{n_{i+1,j}}-\Psi_{n_{i,j}}}{h_i}\left[\frac{k_j}{m_{x_{i,j+1}}}+\frac{k_{j-1}}{m_{x_{i,j}}}
\right]\right]+ \nonumber \\
-\frac{\hbar^2}{4}\left[\frac{\Psi_{n_{i,j+1}}-\Psi_{n_{i,j}}}{k_j}\left[\frac{h_i}{m_{y_{i,j+1}}}+\frac{h_{i-1}}{m_{y_{i-1,j+1}}}
\right]\right]+ \nonumber \\
+\frac{\hbar^2}{4}\left[\frac{\Psi_{n_{i,j}}-\Psi_{n_{i-1,j}}}{h_{i-1}}\left[\frac{k_j}{m_{x_{i-1,j+1}}}+\frac{k_{j-1}}{m_{x_{i-1,j}}}
\right]\right]+ \nonumber \\
+\frac{\hbar^2}{4}\left[\frac{\Psi_{n_{i,j}}-\Psi_{n_{i,j-1}}}{k_{j-1}}\left[\frac{h_i}{m_{y_{i,j}}}+\frac{h_{i-1}}{m_{y_{i-1,j}}}
-\right] \right]+\nonumber \\ E_{c_{i,j}} \;\; \Psi_{n_{i,j}}
\frac{(h_i+h_{i-1})(k_j+k_{j-1})}{4}= E_n \Psi_{n_{i,j}}
\frac{(h_i+h_{i-1})(k_j+k_{j-1})}{4}
\end{eqnarray}

\subsection{Numerical routines and performance}

The discretized non linear Poisson equation is solved with the
Newton-Raphson (NR)algorithm. The sparse system of linear algebraic
equations of each NR step is solved with the package Y12MAF
\cite{netlib}, which is based on Gaussian elimination

The eigenvalue problem resulting from the discretization of the
Schr\"odinger equation in one dimension is solved with the routine
TQLI \cite{netlib}, while in two dimensions is solved with the method
proposed in Ref. \cite{pala_quantum} that allows to reduce computing
time without significant losses in accuracy, by solving the problem in
the momentum space. The method can be applied to structures with
inhomogeneous effective mass and can easily be extended to the full
band structure.

The computing time on an 1800 MHz Pentium IV CPU strongly depends on
the type of simulation: In the case of quantum confinement in one
direction the CPU running time for a 128x61 grid is 37.14 sec, while
in the case of quantum confinement in two directions, with a 113x148
point grid, the CPU running time is about 95.44 sec. These results
represent the worst case, with no initial guess of the unknown
potential. Finally, in the case of a simulation of ballistic current
the running time is strongly affected by the initial guess of the
potential and is between a few minutes and an hour. The initial guess
is also very important in order to avoid convergence problems of the
algorithm.

\section{Examples of Simulation}

\begin{figure}
\begin{center}
\epsfig{file=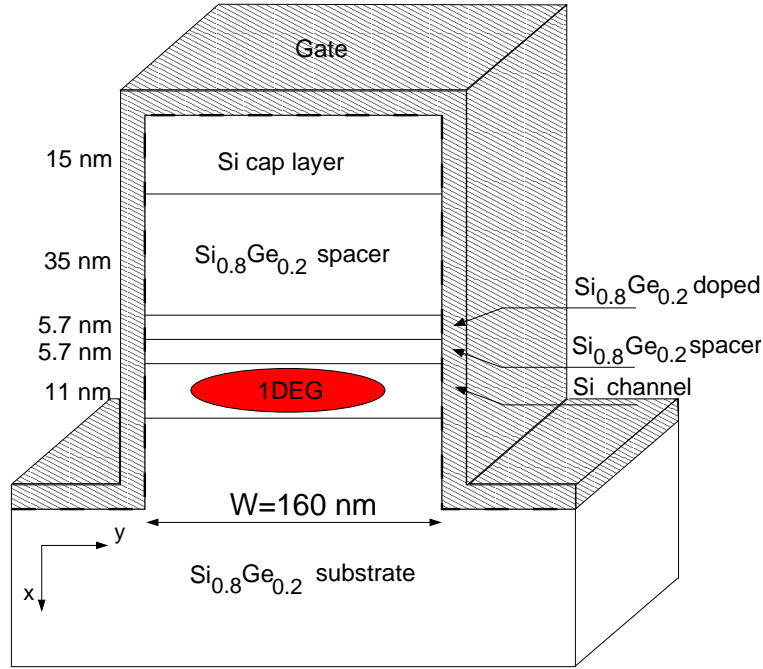,width=10cm}
\caption{Silicon-germanium electron waveguides} \label{sige}
\end{center}
\end{figure}

In this section, two examples of simulations computed on nanoscale
devices are shown. The first structure is a silicon-germanium high
mobility electron waveguide schematically represented in
Figure~\ref{sige}.  It consists of a Si$_{0.8}$Ge$_{0.2}$ virtual
substrate, an 11 nm strained-silicon layer in which the 1DEG forms, a
5.7 nm undoped Si$_{0.8}$Ge$_{0.2}$ spacer layer, a 5.7 nm
Si$_{0.8}$Ge$_{0.2}$ doped layer, with $N_d=10^{18}$ cm $^{-3}$, a 35
nm undoped Si$_{0.8}$Ge$_{0.2}$ spacer and a 15 nm undoped silicon cap
layer. The second spacer is rather thick, in order to prevent the
formation of another electron channel in the silicon cap layer. The
waveguide is 160 nm wide. Finally, we assume that a triple metal gate
is deposited over the structure forming a Schottky contact.  For the
purpose of our simulation, the Schottky junction is reverse-biased and
assumed to be perfectly insulating.

Quantum confinement of carriers in the horizontal ($y$) direction is
provided by selective etching and by the depletion region induced by
acceptor states at the exposed surfaces. This last effect causes the
electrical width of wire to be significantly smaller then the etched
width.  Along the growth ($x$) direction, as a consequence of the band
alignment between strained-silicon and silicon-germanium, a quantum
well for electrons forms. In particular, the strained silicon channel
is grown under a tensile strain and thus two valleys of the conduction
band, along $k_x$, are lowered in energy while the other four valleys
are raised. This condition is required in order to obtain a
confinement region for electrons. In addition, only the two lowest
conduction band valleys are occupied and the energy splitting between
valleys ( 120 meV) leads to strongly suppressed intervalley
scattering.

\begin{figure}
\begin{center}
\epsfig{file=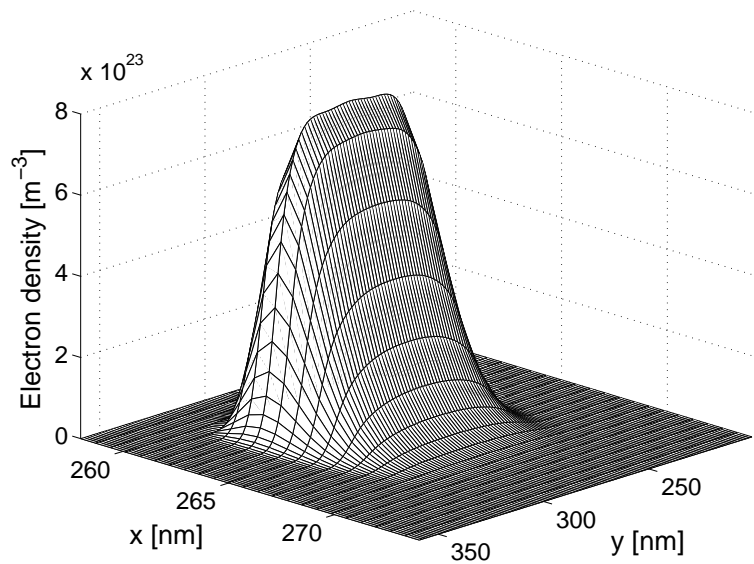,width=10cm}
\caption{Quantum electron density in the strained silicon channel.}
\label{carica}
\end{center}
\end{figure}

\begin{figure}
\begin{center}
\epsfig{file=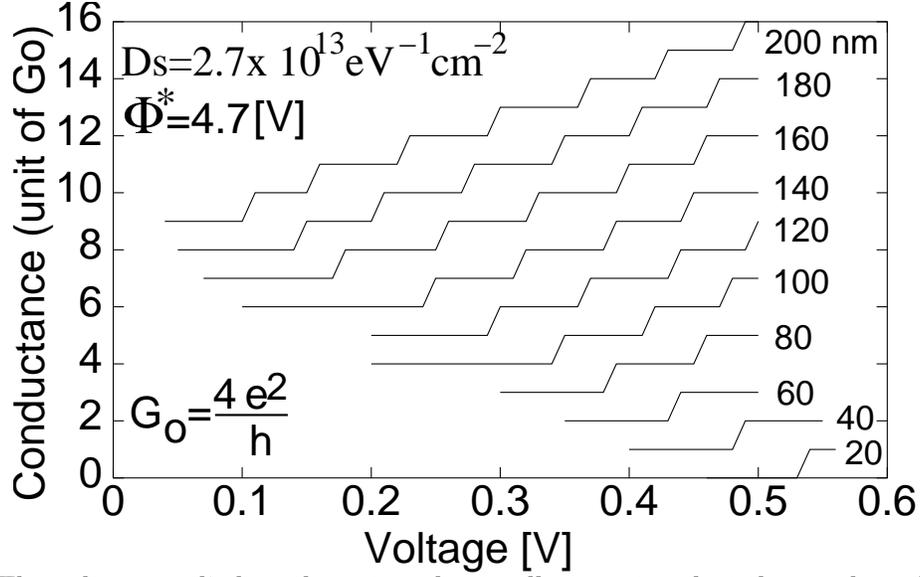,width=12cm}
\caption{The voltage applied on the external gate allows us to select
the number of propagating modes in the waveguide. Hence it's possible
to vary the quantized conductance in the Si channel $G=(2e^2/h)N$ as a
function of gate voltage. N is the number of propagating modes.  For
purpose of presentation each curve is shifted by one conductance
quantum.  } \label{fig4}
\end{center}
\end{figure}

The self-consistent Poisson/Schr\"{o}dinger equation is discretized
onto a nonuniform rectangular grid of $108 \times 137$ points and the
electron concentration has been calculated by solving the
Schr\"{o}dinger equation inside the strained silicon channel. Quantum
electron density is represented in Figure~\ref{carica}, where it is
possible to observe how the electrical waveguide width is about 95 nm,
instead of 160 nm, because of the electron depletion induced by
interface states at the exposed surfaces. By tuning the voltage
applied to the gate, it is possible to vary the electron density in
the channel and hence the number of occupied states. Thus, referring
to the Landauer formula of the quantized conductance $ G=N\left(
2e^2/h \right)$ with N equal to the number of propagating modes, it is
possible to vary, as a function of applied voltage, the quantum
conductance in the channel, as shown in Figure~\ref{fig4}.

\begin{figure}
\begin{center}
\epsfig{file=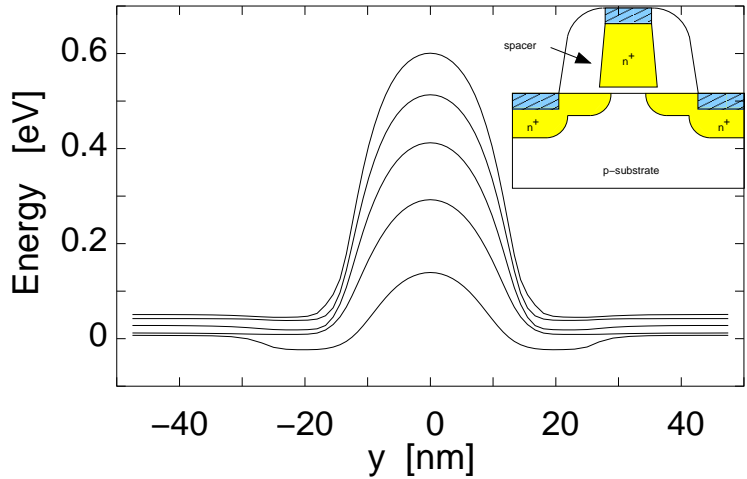,width=10cm}
\caption{Subband energies for a voltage applied to the gate of
1V. Transmission coefficient through the barrier is assumed to be
unity for electrons with energy higher than the peak of the barrier
and zero for electron with energy lower than the peak.}
\label{sub}
\end{center}
\end{figure}

The second simulation example refers to a ``well tempered'' ballistic
MOSFET with channel length of 25nm, proposed by Antoniadis {\em et
al.}~\cite{Antoniadis} and schematically represented in the inset of
Figure~\ref{sub}. The oxide thickness is 1.5 nm and the polysilicon
gate has a donor concentration of $5\times 10^{20}$ cm$^{-3}$. In
order to reduce short channel effects a super-halo doping is implanted
in the channel. The analytic doping profile can be found in
Ref. \cite{Antoniadis}.

Assuming fully ballistic transport within the channel, we have
computed the source-to-drain current. The energy barrier that
electrons encounter traveling from the source towards the drain is
represented in Figure~\ref{sub} for a gate voltage $V_{GS} = 1$~V and
drain-to-source voltage $V_{DS} = 0$~V. Only the first five subbands
are shown. Quantum tunneling has not be considered in our model and
therefore the transmission coefficient is unity above the peak of the
barrier and zero below. Thus, only electrons with energy higher than
the peak can traverse the channel without energy loss and contribute
to the total current.

\begin{figure}
\begin{center}
\epsfig{file=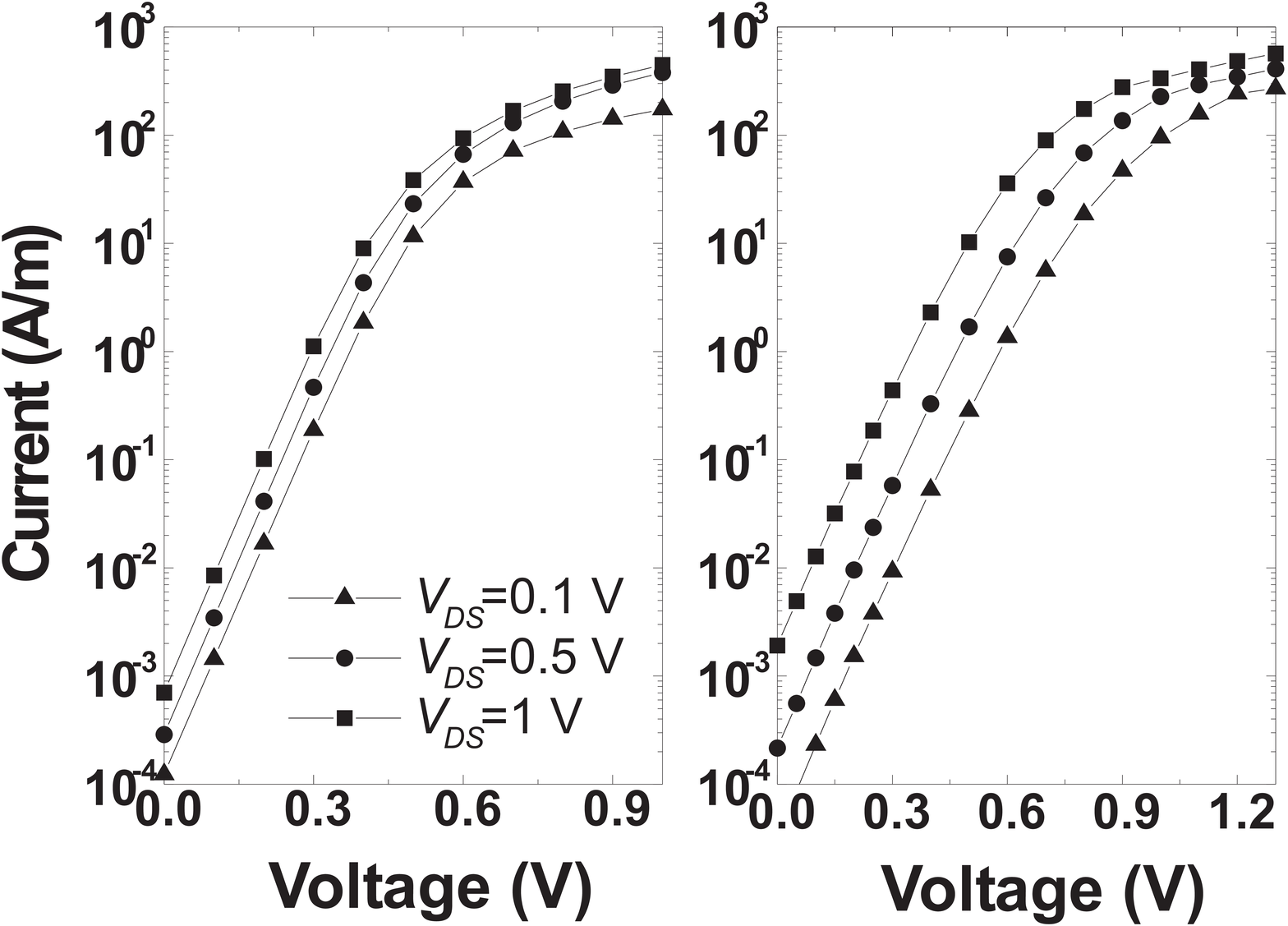,width=10cm}
\caption{Transfercharacteristics of the 25 nm MOSFET
computed with MEDICI (left, from Ref. \protect{Antoniadis})
and with NANOTCAD2D (right).} \label{MEDICI}
\end{center}
\end{figure}

The transfer characteristics obtained with the MEDICI simulator and
with NANOTCAD2D are compared in Figure~\ref{MEDICI}.

\begin{figure}
\begin{center}
\epsfig{file=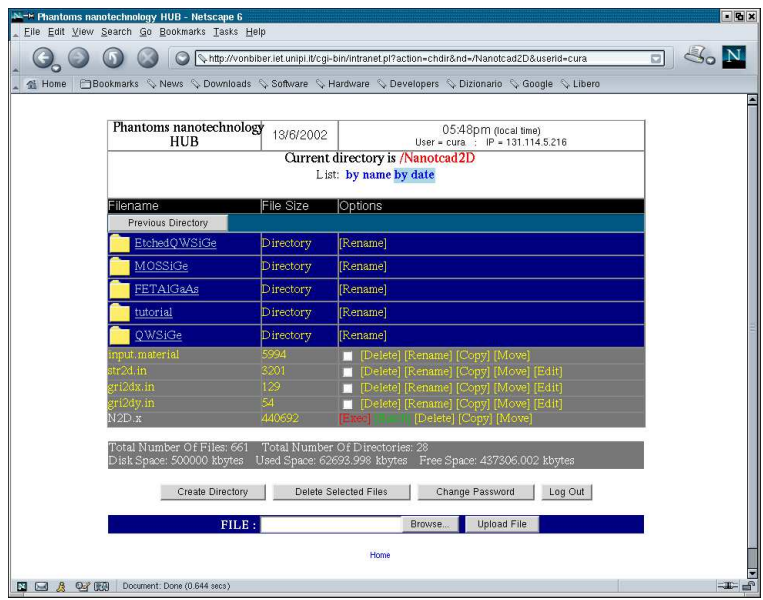,width=10cm}
\caption{Each user can use all the programs available on the
HUB}\label{foto1}
\end{center}
\end{figure}

\begin{figure}
\begin{center}
\epsfig{file=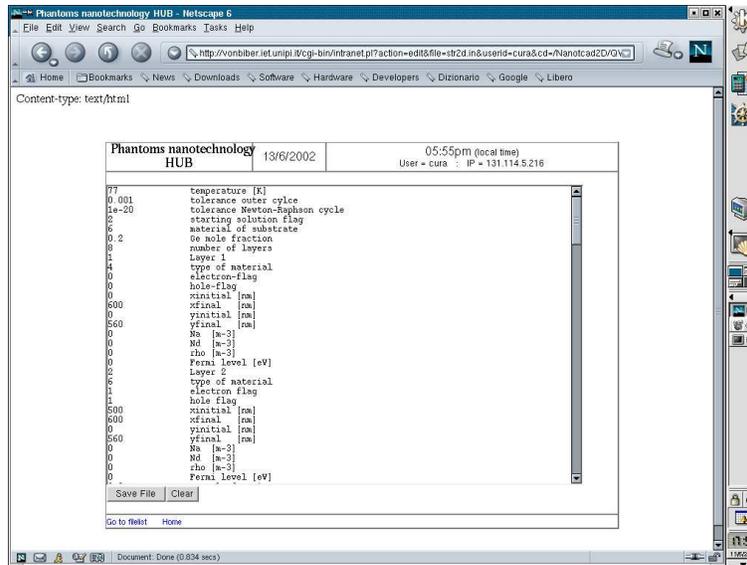,width=10cm}
\caption{The parameters of simulation can be varied in accordance with
the specific user requirements.}\label{foto2}
\end{center}
\end{figure}

The two-dimensional simulator is available, after registration, at the
URL: {\bf{http://www.phantomshub.com}}. In the directory NANOTCAD2D it
is possible to find additional examples of nanoscale semiconductor
devices, among which users can find two different type high mobility
electron waveguides defined by selective etching on a SiGe
heterostructures, a nanoscale ballistic silicon MOSFET and a nanoscale
ballistic AlGaAs field effect transistor~(Figure \ref{foto1}). The
parameters of simulation can be varied according to the specific user
requirements. In particular, input files can be modified by user
directly via a web interface as an HTML form or uploaded from an
external source~(Figure \ref{foto2}). In the same directory it is also
possible to find a detailed tutorial regarding the input data files
structure.

\begin{figure}
\begin{center}
\epsfig{file=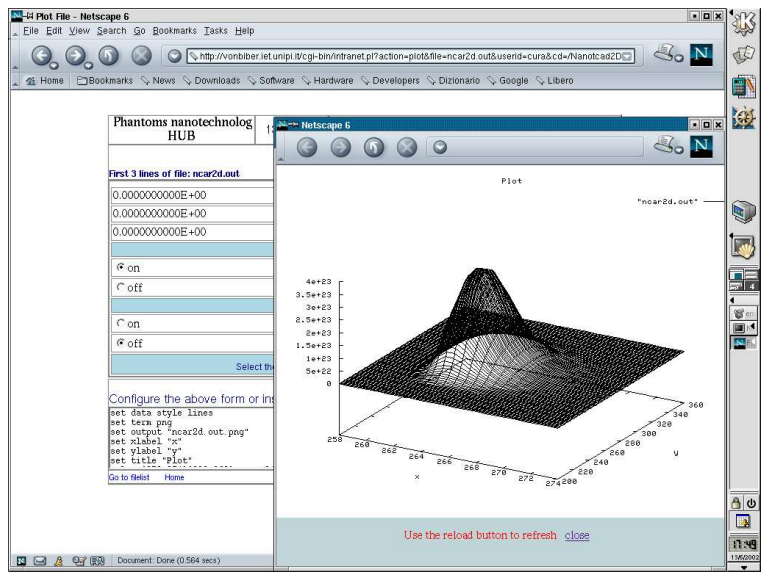,width=10cm}
\caption{The results of simulations can be graphically
represented}\label{foto3}
\end{center}
\end{figure}

Finally, the hub provides basic visualization capabilities for 2D and
3D plots controlled via a web interface~(Figure \ref{foto3}).

\section{Discussion and future developments}

In this paper we have presented a two dimensional quantum simulator
based on the solution of the Poisson/Schr\"odinger equation with the
Box-Integration method. The code solves also the continuity equation
for electrons and holes in the case of ballistic transport, where
propagating states are populated according to the occupation factor of
the originating reservoir. NANOTCAD2D allows to simulate most common
semiconductors, such as silicon, AlGaAs, InGaAs, strained silicon and
silicon germanium and has been successfully used to simulate III-IV
heterostructures, strained-silicon and silicon germanium
heterostructures, CMOS structures. In the case of silicon-germanium
based devices a dedicated procedure allows to take into account the
effects of strain on the energy bands and on band alignment.

The code is presently being improved with full two-dimensional
quantum transport and quantum tunneling, that becomes relevant for
devices with channel length close to 10 nm. In addition, we are
developing a model for quasi-ballistic transport, which accounts
for the possibility that a fraction of carriers undergo elastic
scattering, that would allow a more accurate simulation of
nanoscale field effect transistor at room temperature.

\end{document}